\documentclass[useAMS,usegraphicx,usenatbib]{mn2e}   \usepackage{times}
\usepackage{threeparttable} \usepackage{amssymb}

\newcommand{\scinum}[2]{#1 \times 10^{#2}}

\newcommand{\scinuma}[3]{(#1 \pm #2) \times 10^{#3}}

\newcommand{\decnuma}[2]{#1 \pm #2}

\title[The relativistic iron line and soft excess in Markarian~335]{On
the relativistic  iron line  and soft excess  in the  Seyfert~1 galaxy
Markarian~335\thanks{Based  on observations obtained  with XMM-Newton,
an  ESA science  mission with  instruments and  contributions directly
funded by ESA Member States and NASA}}

\author[P.M.             O'Neill            et~al.]{Paul            M.
O'Neill,$^{1,2}$\thanks{poneill@csu.edu.au} Kirpal
Nandra$^1$\thanks{k.nandra@imperial.ac.uk}, Massimo Cappi$^3$, Anna Lia Longinotti$^4$ and \newauthor Stuart A. Sim$^5$\\
$^1$Astrophysics Group, Imperial  College London, Blackett
Laboratory, Prince Consort Road, London SW7~2AZ\\$^2$School of
Computing \& Mathematics, Charles Sturt University, P.O.
Box~588, Wagga~Wagga NSW 2678, Australia\\$^3$INAF-IASF Bologna, Via Gobetti 101, I-40129 Bologna, Italy\\
$^4$XMM-Newton Science Operation Centre, ESAC, ESA, Apartado
50727, E-28080 Madrid, Spain\\$^5$Max-Planck-Institut fur
Astrophysik, 85741 Garching, Germany}

\begin{document}

\date{Accepted. Received}

\pagerange{\pageref{firstpage}--\pageref{lastpage}} \pubyear{2007}

\maketitle

\label{firstpage}

\begin{abstract}

We report  on a 133~ks  \emph{XMM-Newton} observation of the  Seyfert~1 galaxy
Markarian~335.  The  0.4--12~keV spectrum  contains  an  underlying power  law
continuum,  a soft  excess  below  2~keV, and  a  double-peaked iron  emission
feature  in  the 6--7~keV  range.  We  investigate  the possibility  that  the
double-peaked  emission might  represent the  characteristic signature  of the
accretion disc.  Detailed investigations show that a moderately broad, accretion disc
line  is most likely  present, but  that the  peaks may  be owing  to narrower
components  from more  distant material.  The peaks  at 6.4  and 7~keV  can be
identified, respectively, with the  molecular torus in active galactic nucleus
unification schemes, and  very highly ionized, optically thin  gas filling the
torus. The  X-ray variability  spectra on both  long ($\sim100$~ks)  and short
($\sim1$~ks) timescales  disfavour the recent suggestion that  the soft excess
is an artifact of variable, moderately ionized absorption.

\end{abstract}


\begin{keywords}
galaxies:individual:Markarian~335      --      galaxies:active      --
galaxies:Seyfert -- X-rays:galaxies.
\end{keywords}


\section{INTRODUCTION}

Studying the X-ray emission from active galactc nuclei (AGNs)
offers the potential of observing relativistic effects. The X-ray
emission, being produced deep in the potential well of the
putative black hole, is predicted to be affected by gravitational
redshift and Doppler effects.  Emission lines may thus become
broadened and skewed, the line profiles yielding information on
the inner accretion flow such as the geometry, disc emissivity and
black hole spin \citep[e.g.][]{frs89,s90,l91}.  The Doppler
effects in particular can give rise to a broad Fe~K$\alpha$ line
having a characteristic double-horned structure \citep[][]{chf89}.

At energies below a few keV, an enhancement of flux above the
underlying power law is commmonly seen in AGN spectra.  The origin
of this so-called `soft excess' is a long-standing, unsolved
puzzle in AGN studies \citep[see review by][]{mdp93}. The possible
identification of the soft excess as owing to inverse-Compton
scattering of UV disc photons has been challenged by the apparent
rather constant `temperature' of the excess
\citep[e.g.,][]{gd04,cfg06}. Emission lines associated with disc
reflection \citep[e.g.][]{cfg06}, or the influence of a moderately
ionized `warm' absorber \citep[][]{gd04}, have been proposed
recently to explain the soft excess.

The Seyfert~1 galaxy Markarian~335 ($z = 0.0258$) exhibits both a
soft excess and a broad emission feature in the region of the iron
line. The soft excess has been observed by various missions,
starting with \emph{EXOSAT} \citep[][]{pst87}. A \emph{ROSAT}
spectrum could be modelled equally well as either a double
power-law or a power-law with an absorption edge
\citep[][]{tgm93}. The soft excess in an \emph{ASCA} spectrum
could be modelled as either a blackbody \citep[][]{r97},
additional steep power law component \citep[][]{gtn98} or
reflection from an ionized disc \citep[][]{bif01}. \citet[][]{r97}
found also that the addition of an absorption edge to the
underlying hard power-law component improved the fit. A
\emph{BeppoSAX} spectrum could be modelled similarly to the
\emph{ASCA} spectrum \citep[][]{bmh01}. A narrow iron line,
consistent with arising from neutral material, was detected in the
\emph{ASCA} spectrum with an equivalent width (EW) of $\sim100$~eV
\citep[][]{ngm97}. When fitted instead with a relativistic line
and reflection continuum the EW was $\sim 250$~eV. The
\emph{BeppoSAX} spectrum could likewise be modelled with a
relativistic line having an EW of a few hundred eV. A narrow line
at 6.4~keV is also likely to be present in the \emph{BeppoSAX}
data.

\emph{XMM-Newton} first observed Mrk~335 in 2000 December for
35~ks. \citet[][]{gol02} found excess emission in the 5--7~keV
range above the underlying power-law, which could be described by
relativistically blurred emission from highly ionized iron.  The
combination of reflection from an \emph{unblurred}, highly ionized
disc and Bremstrahlung emission was able broadly to describe the
entire 0.3--10~keV range, with the reflection flux contributing
$\sim 50$~per~cent of the soft excess. \citet[][]{cfg06} also
analysed these data and were able to model the spectrum as
relativistically blurred ionized disc reflection. Importantly,
\citet[][]{gol02} examined the Reflection Grating Spectrometer
(RGS) spectrum and found no evidence of absorption or emission
from ionized gas. \citet[][]{lsn06} recently reanalysed the
European Photon Imaging Camera (EPIC) data from this observation,
confirming the existence of a broad emission feature and noting an
absorption feature at $\sim5.9$~keV.

\emph{XMM-Newton} re-observed Mrk~335 in 2006 for 133~ks. The
purpose of these observations was to better characterise the iron
line emission.  Moreover, being long and nearly uninterrupted,
this observation offers the best data thus far to study the
variability.  We present here an initial analysis of the data
collected by the EPIC PN instrument.


\section{OBSERVATIONS AND ANALYSIS}

\emph{XMM-Newton} observed\footnote{The observation identification
number is 0306870101.} Mrk~335 between 2006 January 03 and 05, for
a duration of 133~ks.  We present here an analysis of the EPIC PN
data.  The observation was conducted in Small Window mode, thus
avoiding photon pile-up, and the Thin filter was used.

Source events were extracted from a circular region, centred at
the X-ray centroid, with a radius of 680 detector pixels
(34\arcsec). Background events were extracted from two rectangular
regions with a combined area 3.4 times larger than the source
region. Background flares were present at the beginning and end of
the observation, and the exclusion of these intervals resulted in
a low-background duration of 115~ks. The final usable data train
contained gaps amounting to $\sim0.3$~per~cent of this duration.

Source and background light curves were extracted for various
energy bands using time resolutions of 200 and 1000~s, and each
time bin was required to be fully exposed. The fractional
root-mean-square (rms) variability spectrum \citep[see,
e.g.][]{vew03} was calculated from the 1000~s light curves, each
of which comprised 104~bins over the 115~ks duration. The
uncertainties in the rms measurements were determined using Monte
Carlo simulations. Each simulation involved perturbing the
observed counting rates with a Gaussian deviate with a standard
deviation equal to the size of the observed error bar. The
fractional rms was then calculated from the synthetic light curve.
We performed 10,000 such simulations, and the standard deviation
of the simulated rms values was adopted as the 1~$\sigma$
uncertainty owing only to Poisson noise. As well measuring variability
in the conventional manner, we also calculated the point-to-point fractional rms
spectrum to probes the variability on relatively short 
timescales \citep[][]{etp02}.

Time-averaged source and background spectra were extracted using
the set of events present in the 1000~s light curves. The spectral
channels were grouped so that: there were no more that 2 groups
per resolution full-width-at-half-maximum, and each group in the
source spectrum possessed at least 20 counts. The source region
spectrum contained $\sim\scinum{1.5}{6}$~counts
(0.4--12~keV\footnote{`PI' channels 380--11995.}), of which
0.14~per~cent are expected to be background. The quoted
uncertainties in the spectral fits correspond to $\Delta\chi^{2} =
1$, unless stated otherwise. These may underestimate the true
uncertainties when multiple parameters are fitted
\citep[][]{lmb76}.


\section{Energy-resolved variability}

\begin{figure}
\rotatebox{0}{\includegraphics[width=0.75\columnwidth]{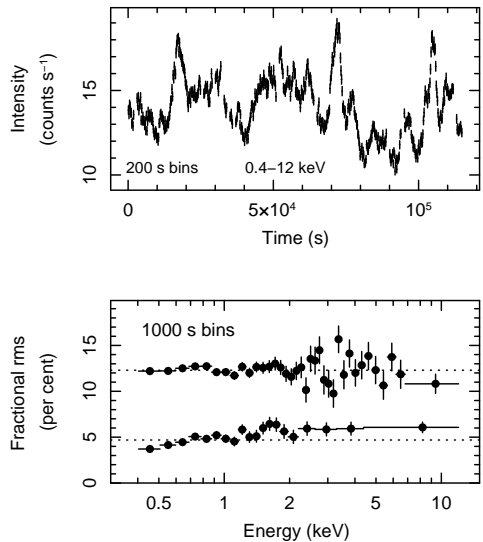}}
\caption{0.4--12~keV light  curve using 200~s bins (top) and fractional rms
  spectra using 1000~s bins (bottom). The upper spectrum was calculated in the
  conventional manner, while the lower shows the point-to-point
  variability. The upper and lower dotted lines show constants of 12.3 and
  4.7~per~cent, respectively.} \label{fig:varplots}
\end{figure}

In Fig.~\ref{fig:varplots}~(top)  we present the 0.4--12~keV light
curve, using a time-resolution of 200~s, which has a fractional
rms variability of $\decnuma{12.24}{0.08}$~per~cent. The
rms spectra, calculated using a time resolution of
1000~s, are shown in Fig.~\ref{fig:varplots}~(bottom). 

The  conventional fractional  rms spectrum  could be  satisfactorily described
with  a constant  of 12.3~per~cent  ($\chi^{2}/\mathrm{DOF} =  34.6/36$).  The
point-to-point  spectrum,  while  still  being  rather flat,  cannot  be  well
described with  a constant ($\chi^{2}/\mathrm{DOF} =  72.3/19$).  Three energy
bands of  importance are 0.4--0.8,  0.8--2 and 2--12~keV: an  ionized absorber
with varying  opacity, if  present, would induce  enhanced variability  in the
0.8--2~keV  band  \citep[][]{gd06}.  Considering  first  the conventional  rms
spectrum,  the  values  in  these  three  bands  are  $\decnuma{12.35}{0.12}$,
$\decnuma{12.32}{0.13}$  and  $\decnuma{12.69}{0.27}$~per~cent,  respectively.
The  uncertainties  in  these  values  suggest that  any  enhancement  in  the
0.8--2~keV range  is constrained to  be less than $\sim0.5$~per~cent.   In the
point-to-point    rms   spectrum    the   corresponding    rms    values   are
$\decnuma{4.23}{0.11}$,               $\decnuma{5.20}{0.12}$               and
$\decnuma{6.12}{0.26}$~per~cent,  which reveal the  absence of  a peak  in the
0.8--2~keV range over short timescales.

The fractional rms  in the 5.6--7~keV band (i.e., the  band containing Fe line
emission; see below) is $\decnuma{12.6}{0.9}$~per~cent. The time-averaged flux
in this band is a factor of 1.12 greater than the underlying power-law. If the
excess of flux in this band  is non-varying, then the rms should be suppressed
by the same  factor. The uncertainty in the  rms is too large to  reach a firm
conclusion regarding the  variability or otherwise of this  feature, but it is
consistent with being as variable as the power-law.


\section{The time-averaged spectrum}

\subsection{A first look}

\begin{figure}
\rotatebox{0}{\includegraphics[width=0.75\columnwidth]{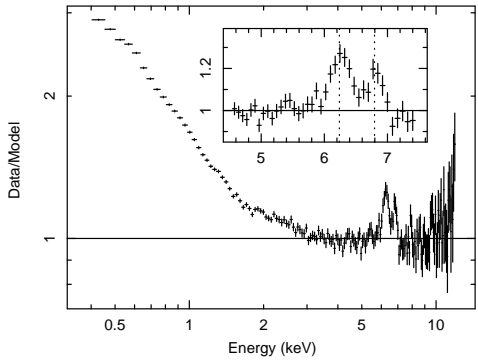}}
\caption{Ratio  between   the  observed  flux   and  the
best-fitting power-law  fitted  over the  observed  3--4.5  and
7.5--12~keV  energy ranges, showing  the iron  line and soft
excess. The inset  shows the double peaked  profile. The energy
scale corresponds to  the observed frame and the left- and
right-hand dotted lines indicate rest energies of 6.4 and
6.97~keV, respectively. The highest  residual seen at 12~keV contributes only $\sim2$
to the  $\chi^2$ of  the power-law fit.} \label{fig:plfit}
\end{figure}

A  power-law  modified by  Galactic  absorption\footnote{The  Galactic NH  was
obtained       using       the       NASA       HEASARC       `nH'       tool,
http://heasarc.gsfc.nasa.gov/docs/tools.html    .}   of    $N_{\mathrm{H}}   =
\scinum{3.99}{20}$~cm$^{-2}$ was  fitted over the observed  frame 3.0--4.5 and
7.5--12~keV energy  ranges. The best-fitting  power-law had a photon  index of
$\Gamma  =  \decnuma{2.00}{0.01}$  with   an  unabsorbed  2--12~keV  flux  and
luminosity\footnote{The values $H_{0} = 70$ and $\lambda_{0} = 0.73$ were used
in            calculating            the            luminosity.}            of
$\scinum{1.75}{-11}$~ergs~cm$^{-2}$~s$^{-1}$                                and
$\scinum{2.66}{43}$~ergs~s$^{-1}$,   respectively.   The  ratio   between  the
observed     flux     and    best-fitting     power-law     is    shown     in
Fig.~\ref{fig:plfit}~(top).  The presence of  a soft excess is clear, reaching
a    maximum   of    $\sim3$   times    the   flux    of    the   extrapolated
power-law.  Double-peaked  line  emission  is  also visible  in  the  6--7~keV
range. Note that the highest  residual seen at 12~keV contributes only $\sim2$
to the  $\chi^2$ of  the power-law fit.  Unlike the  earlier \emph{XMM-Newton}
observation  \citep[][]{lsn06}  there  is   no  clear  absorption  feature  at
$\sim5.9$~keV; we leave a detailed investigation for future work.

We initially attempted to parametrize the line complex using
models with two or more Gaussians fitted over the 3--12~keV range.
A model with two narrow ($\sigma = 1~\mathrm{eV}$) lines yielded
$\chi^{2}/\mathrm{DOF} = 145.4/110$. We allowed the width of the
$\sim6.4$~keV line to vary, and the fit improved to
$\chi^{2}/\mathrm{DOF} = 98.4/109$, with rest-frame line energies
of $\decnuma{6.42}{0.02}$~keV (broad line) and
$\decnuma{7.00}{0.02}$~keV (narrow line). These energies are
consistent with those expected for Fe~{\sc i} K$\alpha$ and
Fe~{\sc xxvi} Ly$\alpha$, respectively. The width of the broad
line was $\sigma = \decnuma{0.19}{0.03}~\mathrm{keV}$, and the EWs
of the broad and narrow lines, respectively, were
$\decnuma{127}{17}$ and $\decnuma{42}{7}$~eV. We then added a
narrow line and the fit improved to $\chi^{2}/\mathrm{DOF} =
95.0/107$. The two narrow lines had best-fitting energies of
$\decnuma{6.40}{0.04}$ and $\decnuma{7.00}{0.03}$~keV, and the
corresponding EWs were, respectively, $\decnuma{20}{9}$ and
$\decnuma{36}{8}$~eV. The broad line had a best-fitting energy,
width and EW of $\decnuma{6.43}{0.05}$, $\decnuma{0.27}{0.06}$~keV
and $\decnuma{115}{14}$~eV, respectively. The joint confidence
intervals of the 6.4~keV line fluxes indicate that these are
greater than zero with a significance of 91 (narrow line) and
$>99.99$~per~cent (broad line).  We then replaced the broad
Gaussian with two narrow lines and the fit worsened to
$\chi^{2}/\mathrm{DOF} = 103.5/106$, with line energies of
$\decnuma{6.28}{0.02}$ and $\decnuma{6.65}{0.04}$~keV.

A double-peaked profile is a natural characteristic of
relativistic disc lines, so it may be possible to interpret the entire
line profile in this framework, with the 6.4~keV peak representing
the red horn and the 7~keV peak the blue horn.  We test this
explicitly below.

\subsection{Neutral disc iron line} \label{sect:ndil}

To  model reflection  from  a neutral  disc  we have  modified the  reflection
continuum model  {\sc pexrav} \citep[][]{mz95}.  The enhanced  model, which we
refer to  as {\sc  pexmon}, includes the  Fe~K$\alpha$ (including  the Compton
shoulder),  Fe~K$\beta$ and Ni~K$\alpha$  emission lines  \citep[see][for more
details]{nog07}.  The dependence of  line flux on photon-index and inclination
is based on  the results of \citet[][]{gf91}.  In  {\sc pexmon} the reflection
continuum  and emission  line fluxes  are  linked, as  expected from  physical
models, and the constraints on  the `reflection fraction' are largely from the
Fe~K$\alpha$ line. To construct a disc  line model we convolved a {\sc pexmon}
component  with the relativistic  blurring model  {\sc kdblur2},  which itself
uses    the     kernel    of    the    {\sc    laor}     disc    line    model
\citep[][]{l91,fvn02}. Following \citet{nog07},  the disc emissivity power-law
index  was fixed  to  break  from a  value  of 0  within  the `break  radius',
$r_{\mathrm{br}}$, to a value of $-3$  for larger radii. The disc inner radius
was fixed at its minimum value of 1.235~r$_{\mathrm{g}}$, which corresponds to
the innermost stable orbit for a  maximally spinning black hole, and the outer
radius was fixed at the maximum permitted model value of 400~r$_{\mathrm{g}}$.
The  inclination and  $r_{\mathrm{br}}$ were  free to  vary. With  the assumed
emissivity, this model  well approximates a point source  illuminating a slab,
with  a  peak  emissivity  at  $r_{\mathrm{br}}$  and a  height  of  the  same
order. Note that this emissivity law is non-relativistic, and does not account
for   enhanced  inner  disc   reflection  owing   to  light   bending  effects
\citep[e.g.][]{mkm00}.

A single neutral disc line was unable to model the entire line
profile. A large component of line flux is required at the Fe~{\sc
i} rest energy of 6.4~keV.  This would require the red-horn to be
only slightly shifted in energy via gravitational and Doppler
effects. It is not possible to then also produce a blue horn that
is significantly shifted from 6.4~keV. Moreover, the blue horn is
expected to be more intense than the red horn.

We therefore  attempted to  model the profile  with a combination  of distant,
neutral  reflection and  a disc  line.  The  former component  is  intended to
account primarily for the narrow emission  at 6.4~keV and we modelled it using
an unblurred  {\sc pexmon} component  with a fixed inclination  of 60~degrees.
The disc line, on the other hand, is intended to model the blue wing and broad
component   of  the   profile.   This   model  yielded   a   satisfactory  fit
($\chi^{2}/\mathrm{DOF} = 99.0/110$).  The  break radius was constrained to be
less than  70~r$_{\mathrm{g}}$ (95~per~confidence), with  a best-fitting value
of 4~r$_{\mathrm{g}}$.  While the overall  shape of the profile is reproduced,
the blue-horn  in this  model is unable  to account  for the sharpness  of the
observed line at 7~keV.

We next  explored the  possibility that there  is an Fe~{\sc  xxvi} Ly$\alpha$
emission component contributing  to the blue wing. This  line might physically
be  produced via  fluorescence in  optically thin  plasma. We  added  a narrow
($\sigma = 1$~eV) Gaussian with a fixed energy of 6.97~keV to the best-fitting
disc line model. The  fit was significantly improved ($\chi^{2}/\mathrm{DOF} =
89.2/109$),  with a  reduction  in $\chi^{2}$  of  9.8. The  break radius  and
inclination were poorly constrained, with the one-sided 95~per~cent confidence
intervals ($\Delta  \chi^{2} = 4.61$) restricting the  radius and inclination,
respectively,  to $>15$~r$_{\mathrm{g}}$  and  $>22$~degrees. Indeed,  varying
$r_{\mathrm{br}}$  through values  above about  $\sim$150~r$_{\mathrm{g}}$ did
not yield any variation in $\chi^{2}$. Relative to the power-law continuum, the
reflection   fraction   of   the    blurred   {\sc   pexmon}   component   was
$\decnuma{0.6}{0.1}$, which  corresponds to the disc subtending  a solid angle
of $1.2\pi$ for a slab geometry. The reflection fraction of the unblurred {\sc
pexmon}  component  and the  EW  of the  Fe~{\sc  xxvi}  Ly$\alpha$ line  were
$\decnuma{0.31}{0.08}$  and $\decnuma{34}{7}$~eV,  respectively, and  the line
flux was $\scinuma{4.7}{1.0}{-6}$~photons~cm$^{-2}$~s$^{-1}$.

Finally, we removed the {\sc pexmon} component representing distant reflection
to investigate whether the profile  could be modelled with only a relativistic
disc  line and  a narrow  Fe~{\sc  xxvi} Ly$\alpha$  line.  In  this case  the
best-fitting     break    radius     was    constrained     to     be    above
$\sim$100~r$_{\mathrm{g}}$,   and  the   reflection   fraction  increased   to
$\decnuma{0.7}{0.1}$.   The fit  became significantly  worse,  with $\chi^{2}$
increasing by 7.4.

\subsection{Ionized disc iron line}

Part of the difficulty in reproducing the observed profile with a neutral disc
line is  that, in order to  reproduce the sharp blue  peak at $\sim  7$ keV, a
relatively   high  inclination  with   minimal  gravitational   broadening  is
required. However, the \emph{red horn} in such a model, being narrow and at an
energy  below 6.4~keV,  fails  to fit  the  broad observed  excess around  the
Fe~{\sc i}  rest energy.  An ionized disc,  on the other hand,  can produce Fe
lines  with rest  energies between  6.4 and  7~keV.  With  variable  $\xi$ and
inclination an ionized disc model  might, then, be flexible enough to describe
the observed profile.

To   construct   an    ionized   disc   line   model   we    used   the   {\sc
reflion}\footnote{http://heasarc.gsfc.nasa.gov/xanadu/xspec/models/reflion.html}
model  \citep[][]{rf05}, which  incorporates both  line emission  with Compton
broadening and the reflection continuum. {\sc reflion} is defined at a limited
number  of   values  of  ionization   parameter,  $\xi$,  and   $\Gamma$,  and
interpolation between  these grid  points may yield  deviations from  the true
model spectrum  (2006, R. Ross, priv. comm.).  We consider $\xi$ to  be a free
parameter but keep its value fixed during minimisation. 

We  constructed a  model comprising  distant reflection  ({\sc  pexmon}), plus
relativistically blurred ionized reflection ({\sc kdblur2} and {\sc reflion}),
plus a  narrow Fe~{\sc  xxvi} Ly$\alpha$ line.   The ionization  parameter was
fixed  at  the   minimum  permitted  value  of  $\xi   =  30$,  which  yielded
$\chi^{2}/\mathrm{DOF}  =  87.4/108$. We  then  increased  $\xi$  to its  next
defined value of  $\xi = 100$, and the best-fitting  model yielded $\chi^{2} =
87.8/108$.  When the ionization was increased further, to $\xi = 300$, the fit
worsened to $\chi^{2} = 99.2/108$. An ionized disc model is thus not preferred
over a  neutral disc. We  repeated these three  fits without a narrow  line at
6.97~keV, and confirmed that, as for  the neutral disc model, the inclusion of
the Fe~{\sc xxvi} Ly$\alpha$ line improved the fit.

\subsection{Partial covering absorber}

The broad emission feature in the earlier \emph{XMM-Newton}
spectrum could be well-described by either a relativistic disc
line (the preferred interpretation) or a neutral partial covering
absorber model \citep[][]{lsn06}. One of the goals of obtaining
the recent, longer observation was to break the degeneracy between
these two interpretations. We have, therefore, attempted to fit
the new data with a partial covering model.

We initially fitted the spectrum with a partial covering absorber
plus a {\sc pexmon} component to represent distant, neutral
reflection, which yielded $\chi^{2}/\mathrm{DOF} = 101.9/111$. The
addition of a narrow Fe~{\sc xxvi} Ly$\alpha$ emission line
improved the fit ($\chi^{2}/\mathrm{DOF} = 96.0/110$), with a
covering fraction of $\decnuma{36}{6}$~per~cent. This model, while
yielding a worse fit compared to the neutral disc line, is
formally acceptable, so we cannot rule out a partial covering
interpretation. The best-fitting photon index for this model is
$\Gamma \sim 2.3$, compared to $\Gamma \sim 2.1$ for the neutral
disc line model.

\subsection{The soft excess}

The {\sc reflion} model includes the soft X-ray line emission that
is proposed by, e.g., \citet[][]{cfg06} to explain the soft
excess. Detailed modelling of the soft excess is beyond the scope
of this Letter. However, to determine roughly the extent to which
the soft excess might be attributed to ionized reflection we
examined the ratio between the observed spectrum in the 0.4--3~keV
range and the extrapolated best-fitting 3--10~keV ionized disc
model.

We  initially  used the  best-fitting  model  with $\xi  =  30$  and Solar  Fe
abundance. The  observed flux at 0.4~keV  is $\sim2.2$ times  greater than the
extrapolated  model. \citet[][]{cfg06}  fitted  the earlier  \emph{XMM-Newton}
observation with an  Fe abundance of 0.7.  We  therefore reduced the abundance
to  0.5,  which is  the  closest  value to  0.7  for  which  {\sc reflion}  is
defined. This reduces the Fe line  flux relative to the soft flux. We refitted
the 3--10~keV  spectrum and the  0.4--3~keV residuals were  up to a  factor of
$\sim1.6$  above the  extrapolated model.  Reducing  the abundance  to the  next
lowest  defined value  of  0.2  yielded a  significantly  worse 3--12~keV  fit
($\Delta \chi^{2} = 19.5$).

We  then  increased the  ionization  parameter  to $\xi  =  300$,  with an  Fe
abundance  of  0.5.  This  increased  the  model soft  flux  and  reduced  the
0.2--3~keV  residuals  to be  up  to  a factor  of  only  $\sim1.1$ above  the
model. However, the rather smooth shape  of the soft excess below 2~keV is not
reproduced. An increase in the  relativistic blurring could produce a smoother
spectrum, but in  our simple test the disc line  parameters are constrained by
the requirement to fit the 3--12~keV spectrum.


\section{DISCUSSION}

We have conducted an initial analysis of a 133~ks
\emph{XMM-Newton} observation of the Seyfert~1 galaxy
Markarian~335. The time-averaged spectrum in the 3--12~keV range could be
described as a power law continuum ($\Gamma =
\decnuma{2.00}{0.01}$) with a double peaked emission feature in
the 6--7~keV range.  A strong soft excess is present below
$\sim3$~keV.

\subsection{The Fe K$\alpha$ emission}

A relativistic disc  line alone could not describe the  line profile.  A model
comprising a disc line and reflection from neutral, distant material yielded a
satisfactory fit. The  blue horn in the best-fitting disc line  is not able to
describe very  well the sharpness of  the observed peak.   Increasing the disc
ionization did  not improve the fit.   However, the addition  of Fe~{\sc xxvi}
Ly$\alpha$ emission did  significantly improve the fit. The  iron line profile
is, then, likely  owing to a superposition of  distant reflection from neutral
material, relativistically  blurred reflection  from a neutral  accretion disc
and a  narrow emission line from highly  ionized gas.  Note that  the level of
relativistic blurring in our fit is rather moderate. Therefore, our physically
motivated emissivity  law (see Section~\ref{sect:ndil}) is  preferred over the
often-used \emph{unbroken} powerlaw.

We identify the distant reflection  component as originating from the torus in
AGN unification schemes. For  the torus geometry defined in \citet[][]{ghm94},
an assumed {\sc pexmon} inclination angle of 60~degrees corresponds to a torus
half-opening angle  of $\sim$30~degrees.  The observed  {\sc pexmon} component
has a  reflection fraction of  $\decnuma{0.31}{0.08}$, compared to a  value of
$\sim$0.9 expected for the assumed  opening angle. This discrepancy is perhaps
explained  by the opening  angle being  greater than  assumed. A  half opening
angle of  $\sim75$~degrees yields a  reflection fraction compatible  with that
observed.   The reflection  flux depends  on the  optical depth  on  the torus
\citep[][]{ghm94},  and this might  also explain  the low  observed reflection
fraction.   The reflection  fraction of  both the  distant and  the  disc line
components  are well  within the  ranges  observed in  other Seyfert  galaxies
\citep[e.g.][]{nog07}.

The Fe~{\sc xxvi} Ly$\alpha$ emission  line, being narrow, must originate from
optically thin material.  An ionized  disc, for example, cannot produce such a
narrow line, owing to Compton broadening.  The thin gas might be identified as
the hot gas  that fills the torus and which is  responsible for scattering the
broad line optical  emission into the line of  sight \citep[][]{kk87,am85}. An
Fe~{\sc xxvi} absorption  edge at 9.29~keV should accompany  the Fe~{\sc xxvi}
Ly$\alpha$  line.  For  a fluorescent  yield of  0.7, the  upper limit  on the
Fe~{\sc xxvi} absorption edge flux in the spectrum, together with the observed
line flux,  corresponds to  a 95~per~cent \emph{lower}  limit on  the covering
fraction  of $\sim$100~per~cent.  This  rather large  fraction can  perhaps be
explained if  the torus does not  obscure the hot gas  on the far  side of the
nucleus, and there may be a deficit of gas in the line-of-sight to
the central engine.

A partial covering interpretation for the line profile was
satisfactory, yet worse than disc reflection. Partial covering
yields a steeper photon index than for the disc line, so high
energy data (e.g., \emph{Suzaku}) will be able to robustly
distinguish between these two interpretations \citep[see,
e.g.][]{rwb04}.

\subsection{The soft excess}

The soft  excess seen  in many AGN  X-ray spectra  is possibly an  artifact of
ionized  absorption.   The  RGS   data  from  the  previous  \emph{XMM-Newton}
observation of  Mrk~335 \citep[][]{gol02},  however, revealed no  evidence for
absorption or emission  from ionized gas.  We visually  inspected the combined
fluxed RGS spectrum from the new data  and the only clear feature is an O~{\sc
i} absorption line at 0.54~keV. However,  this lack of lines is possibly owing
to strong Doppler smearing \citep[][]{gd04}.

A moderately ionized absorber can yield  a soft excess of the kind observed in
Mrk 335, where  absorption by O{\sc vii}, O{\sc viii}  and Fe-L are important.
For lower  $\xi$ the  soft spectrum would  show heavy absorption,  whereas for
very high ionization no significant effects on the soft spectrum would be seen
at all.  For the absorbers proposed to explain soft excesses, \citet[][]{gd06}
argued  that variations in  the continuum  luminosity will  necessarily induce
variations  in  ionization,  and  they  predicted  excess  variability  around
1~keV. Therefore,  the presence of an  absorber might be  inferred by enhanced
variability in the 0.8--2~keV range.  Mrk~335 exhibits no such enhancement, so
we cannot invoke  a moderately ionized absorber with  varying $\xi$ to explain
its soft excess. It is worth noting that a flat rms spectrum might be produced
either in the presence of \emph{non-varying} absorption or in the context of a
partial    covering    absorber   model    \citep[see][for    the   case    of
1H~0707$-$495]{bfs02,tbg04}.

Perhaps, then, the excess is owing to disc reflection.  \citet[][]{cfg06} used
an  iron  abundance  of  0.7  were  able to  satisfactorily  fit  the  earlier
\emph{XMM-Newton} observation of Mrk~335.  The  purpose of our simple test was
to investigate the extent to which  the disc line model could explain the soft
excess. When  we extrapolated  the best-fitting 3--12~keV  model down  to soft
energies we found  that disc reflection underpredicts the  soft excess flux by
up  to a  factor  of  $\sim1.6$, even  with  a reduced  Fe  abundance of  0.5.
Increasing the  disc ionization reduced  the residuals, but the  observed soft
excess  continuum  is  much  smoother  than the  model.   A  more  complicated
reflection model could  perhaps be invoked.  For example,  the soft flux might
originate from very close to the black  hole, in a region that is both ionized
and highly smeared, while the Fe line flux is owing to neutral reflection from
further out in the disc. We note  in passing that a flat rms spectrum might be
associated   with  `regime  I'   in  the   so-called  `light   bending  model'
\citep[][]{mf04}. However,  the relatively  large source height  inferred from
our disc  line modelling,  and the low  reflection flux, is  inconsistent with
this regime.


\section*{Acknowledgments}

The  authors acknowledge financial  support from  PPARC (PMO)  and
the Leverhulme trust (KN). We thank Andy Fabian for the
relativistic blurring code and Randy Ross for assistance with {\sc
reflion}.


\bibliographystyle{mn2e} 

\bibliography{oneill_mkn335}


\label{lastpage}

\end{document}